\documentstyle[12pt]{article}
\begin{document}
\begin{flushright}
HUB-EP-97/82
\end{flushright}
\medskip
\renewcommand{\thefootnote}{\fnsymbol{footnote}}
{\Large\bf Relativistic quark model approach to 
weak hadronic decays of $B$ mesons\footnotemark[1]} 
\\[2mm]
{\it D. Ebert,$^a$\footnotemark[2]
\underline{R. N. Faustov},$^b$\footnotemark[3]{} 
V. O. Galkin $^{a,b}$\footnotemark[3]}\\
[2mm]
{\small$^a$ Institut f\"ur Physik, Humboldt--Universit\"at zu Berlin,
Invalidenstr.110, D-10115 Berlin, Germany}\\
{\small$^b$ Russian Academy of Sciences, Scientific Council for
Cybernetics, Vavilov Street 40, Moscow 117333, Russia}


\setcounter{footnote}{1}
\footnotetext{ Talk given at the
IV International Workshop on Progress in Heavy Quark Physics,
Rostock, 20-22 September 1997.}
\setcounter{footnote}{2}
\footnotetext{Supported in part by {\it Deutsche
Forschungsgemeinschaft} under contract Eb 139/1-3.}
\setcounter{footnote}{3}
\footnotetext{Supported in part by {\it Russian Foundation 
for Fundamental Research}
 under Grant No.\ 96-02-17171.}
\renewcommand{\thefootnote}{\arabic{footnote}}
\setcounter{footnote}{0}

\begin{abstract}
\noindent
The energetic exclusive two-body nonleptonic decays of $B$ mesons are
investigated in the framework of the relativistic quark model within
the factorization approximation. 

\end{abstract}

The  investigation of exclusive nonleptonic decays of $B$ mesons
represents an important and complicated theoretical problem. In
contrast to the exclusive semileptonic decays, where  the weak current
matrix elements between meson states  are involved, nonleptonic decays
require the evaluation of  hadronic matrix elements of the local
four-quark operators. To simplify the analysis it is usually assumed
that the matrix element of the current-current weak interaction
factorizes into the product of two single current matrix elements.
Thus the problem reduces to the calculation of the meson form
factors, parametrizing the hadronic matrix elements of weak currents as
in the case of semileptonic decays, and the meson decay constants,
describing leptonic decays \cite{BSW}.
Factorization can be intuitively justified for the energetic
nonleptonic decays \cite{JB}. In these decays the final hadrons are
produced in the form of point-like colour-singlet objects with a large
relative momentum. And thus the hadronization of the decay products
occurs  after they are too far away for strongly interacting with
each other, providing the possibility to avoid  final state
interactions.

The factorization approach to two-body nonleptonic decays implies 
that the decay amplitude
can be approximated by the product of one-particle matrix elements. As an
example, for the decays $B^0\to D^{+(0)}\pi^{-(0)}$ we have:
\begin{eqnarray}
&&\label{factor} \langle D^+\pi^-|H_{eff}|B^0\rangle= \frac{G_F}{\sqrt{2}}
V_{cb}V_{ud}a_1(\mu)\langle D^+|(\bar
cb)_{V-A}|B^0\rangle\langle \pi^-|(\bar du)_{V-A}|0\rangle,\cr 
&&\langle D^0\pi^0|H_{eff}|B^0\rangle= \frac{G_F}{\sqrt{2}}
V_{ub}V_{cd}a_2(\mu)\langle D^0|(\bar
cu)_{V-A}|0\rangle\langle \pi^0|(\bar db)_{V-A}|B^0\rangle,
\end{eqnarray}
where \cite{NS}
\begin{equation}
\label{amu} a_1(\mu)=c_1(\mu)+\zeta c_2(\mu), \qquad
a_2(\mu)=c_2(\mu) +\zeta c_1(\mu),
\end{equation}
$c_{1,2}$ are Wilson coefficients and $\zeta$ is 
a hadronic parameter of order $1/N_c$, which 
accounts also for nonfactorizable corrections (such as e.g. colour octet 
contributions) and therefore is process dependent \cite{NS}. However, as 
it is argued in \cite{NS} for most of two-body  $B$ decays  
the variation of $\zeta$ is rather small and, to a good approximation,
this process dependence can be discarded.

The matrix elements of the weak current  between meson states
or  the vacuum and a meson state are parametrized by the form factors 
$F_{1,0}(q^2)$, $A_{0,1,2}(q^2)$, $V(q^2)$ and 
decay constants $f_{P,V}$ \cite{BSW}.
All these form factors have been calculated  in the framework of the
relativistic quark model, based on the quasipotential approach 
\cite{8,semil}. 

For the  case of the heavy-to-heavy ($B\to D^{(*)}$) 
meson decays our model form factors satisfy
all the constraints imposed  by heavy quark
symmetries. We have determined the Isgur-Wise function and
subleading form factors as well as the second order
power corrections at the point of zero recoil of final meson \cite{8}.
The obtained values of $B\to D^{(*)}$ form factors at $q^2=0$ are
\cite{EFG}
\begin{eqnarray}
&&F_1(0)=F_0(0)=0.63, \quad V(0)=0.79,\cr
&&A_0(0)=0.63, \quad A_1(0)=0.62,\quad A_2(0)=0.61.
\end{eqnarray}

In the case of heavy-to-light $B\to\pi(\rho)$ decays 
we have calculated corresponding
form factors at the point of $q^2=0$ using the expansion in inverse 
powers of the heavy $b$-quark mass and large recoil momentum of light
meson (of order $\sim m_b/2$) up to second order \cite{semil}.  
The found values of $B\to\pi(\rho)$ form factors at the point of
maximum recoil ($q^2=0$) are \cite{EFG}
\begin{eqnarray}
\label{lff}
&& F_1(0)=F_0(0)=0.21, \quad V(0)=0.29,  \cr
&& A_0(0)=0.18,   \quad A_1(0)=0.27, \quad A_2(0)= 0.30.
\end{eqnarray}
We have also determined 
the $q^2$ behaviour of the heavy-to-light form factors.
It is important to note that the form factors $A_1$ and $F_0$ in
our model have a different $q^2$ dependence than the
other form factors \cite{semil,EFG}. 

Let us apply these form factors to the calculation of nonleptonic
decays.
In the factorization approximation one can distinguish three classes
of $B$ meson nonleptonic decays \cite{BSW}: the `class I'
transitions, such as $\bar B^0\to M_1^+ M_2^-$, where only the term
with $a_1$  contributes (i.e. both mesons are
produced by charged currents); `class II' transitions, such as $\bar
B^0\to M_1^0 M_2^0$, where only the term with $a_2$ 
contributes (i.e. both mesons are produced by neutral currents) and
`class III' transitions, such as $B^-\to M_1^0 M_2^-$, where both
terms can contribute coherently.

The results of the calculation of the nonleptonic branching ratios for
$B\to D^{(*)}$ and $B\to \pi(\rho)$ transitions are given in Tables~1
and 2 in comparison with other model predictions
\cite{NRSX}-\cite{Ch} and experimental data. The $B\to
D^{(*)}D_s^{(*)}$ decay branching ratios are presented for
completeness.

\begin{table*}[hbt]
\caption{Predicted branching ratios for  $B\to D^{(*)}M$ nonleptonic
decays in terms of $a_1$ and $a_2$
(in \%). Our model branching ratios are quoted for values of
$a_1=1.05$ and $a_2=0.25$ in comparison with experimental data (in
\%).  We use the values  $|V_{cb}|=0.038$ and $f_D=f_{D^*}=0.220$ GeV,
$f_{D_s}= f_{D_s^*}=0.260$ GeV, $f_{a_1}=0.205$ GeV [7] for our
estimates.}
\label{table:nonl1}
\small
\begin{tabular}{lcccc}
\hline
\hline
Decay & our result & \cite{NRSX}& our result& experiment \cite{BHP}\\
\hline
$\bar B^0\to D^+\pi^-$ & $0.29 a_1^2$ & $0.264 a_1^2$
& 0.32 & $0.31\pm0.04\pm0.02$\\
$\bar B^0\to D^+\rho^-$ & $0.79 a_1^2$ & $0.621 a_1^2$
& 0.87 & $0.84\pm0.16\pm0.05$\\
$\bar B^0\to D^{*+}\pi^-$ &$ 0.26 a_1^2$ & $0.254 a_1^2$
& 0.28 & $0.28\pm0.04\pm0.01$ \\
$\bar B^0\to D^{*+}\rho^-$ &$ 0.81 a_1^2$ & $0.702 a_1^2$
& 0.88 & $0.73\pm0.15\pm0.03$ \\
$\bar B^0\to D^+a_1^-$ & $0.78a_1^2$ & $0.673a_1^2$
& 0.86 & $0.83\pm0.20\pm0.08$  \\
$\bar B^0\to D^{*+}a_1^-$ & $0.92a_1^2$ & $0.970a_1^2$
& 1.02 & $1.16\pm0.25\pm0.05$\\
$\bar B^0\to D^{+}D_s$ & $1.37a_1^2$ &$1.213a_1^2$
& 1.51& $0.74\pm0.22\pm0.18$\\
$\bar B^0 \to D^+D_s^{*-}$ & $0.685 a_1^2$ & $0.859a_1^2$
& 0.75 & $1.14\pm0.42\pm0.28$ \\
$\bar B^0\to D^{*+}D_s^-$ & $0.82 a_1^2$ & $0.824a_1^2$
& 0.90 & $0.94\pm0.24\pm0.23$ \\
$\bar B^0\to D^{*+}D_s^{*-}$ & $2.50a_1^2$ & $2.203a_1^2$
& 2.75 & $2.00\pm0.54\pm0.49$ \\
$\bar B^0\to D^0\pi^0$ & $0.058a_2^2$ & $0.20a_2^2$
& 0.0036 & $<0.012$ \\
$\bar B^0 \to D^{*0}\pi^0$ & $0.056a_2^2$ & $0.21a_2^2$
& 0.0035 & $<0.044$ \\
$\bar B^0 \to D^0\rho^0$ & $0.053a_2^2$ & $0.14a_2^2$
& 0.0033 & $<0.039$ \\
$\bar B^0 \to D^{*0}\rho^0$ & $0.156a_2^2$ & $0.22a_2^2$
& 0.0098 & $<0.056$ \\
$B^-\to D^0\pi^-$ & $0.29(a_1+0.64a_2)^2$ & $0.265(a_1+1.230a_2)^2$
& 0.43 & $0.43\pm0.05\pm0.03$ \\
$B^-\to D^{*0}\pi^-$ & $0.27(a_1+0.69a_2)^2$ &$0.255(a_1+1.292a_2)^2$
& 0.40 & $0.39\pm0.06\pm0.03$ \\
$B^-\to D^0\rho^-$ & $ 0.81(a_1+0.36a_2)^2$ & $0.622(a_1+0.662a_2)^2$
& 1.06 & $0.92\pm0.14\pm0.03$ \\
$B^-\to D^{*0}\rho^-$ & $0.83(a_1^2+0.39a_2^2$ &
$0.703(a_1^2+0.635a_2^2$ & 1.17& $1.28\pm0.20\pm0.09$ \\
& $\phantom{0.83}+1.15a_1a_2)$ &$\phantom{0.703}+1.487a_1a_2)$ & &\\
$B^-\to D^0D_s^-$ & $1.40a_1^2$ & $1.215a_1^2$
& 1.55 & $1.36\pm0.28\pm0.33$ \\
$B^-\to D^0D_s^{*-}$ & $0.70a_1^2$ & $0.862a_1^2$
& 0.77 &$0.94\pm0.31\pm0.23$ \\
$B^-\to D^{*0}D_s^-$ & $0.84a_1^2$  & $0.828a_1^2$
& 0.92 & $1.18\pm0.36\pm0.29$ \\
$B^-\to D^{*0}D_s^{*-}$ & $2.56a_1^2$ & $2.206a_1^2$
& 2.80 & $2.70\pm0.81\pm0.66$ \\
\hline
\hline

\end{tabular}
\end{table*}

\begin{table*}[hbt]
\small
\caption{Predicted branching ratios for $B\to\pi(\rho)M$ nonleptonic
decays. We use the experimental values for $f_\pi$, $f_\rho$
[7] and the value of $|V_{ub}|=0.0052$ [5] for our
model estimates. All numbers are branching ratios $\times 10^5$.}
\label{table:nonl2}
\begin{tabular}{lccccc}
\hline
\hline
Decay & our result & our result & \cite{Dean}&\cite{Ch}&
experiment UL \cite{PDG,BHP}\\
\hline
$\bar B^0\to\pi^+\pi^-$ & $0.331|V_{ub}|^2a_1^2$
& 0.99 & 1.8& & $<1.5$ \\
$\bar B^0\to\pi^+\rho^-$ & $0.857|V_{ub}|^2a_1^2$
& 2.55 & 4.8 &  & \\
$\bar B^0\to\rho^+\pi^-$ & $0.234|V_{ub}|^2a_1^2$
& 0.70 & 0.4 &  &  \\
$\bar B^0\to\pi^{\pm}\rho^{\mp}$ &$1.09|V_{ub}|^2a_1^2$
& 3.25& 5.2& & $<8.8$ \\
$\bar B^0\to\rho^+\rho^-$ & $ 0.794|V_{ub}|^2a_1^2$
& 2.36 & 1.3 & & $<220$\\
$\bar B^0\to\pi^0\pi^0$ & $ 0.17|V_{ub}|^2a_2^2$
& 0.028 & 0.06 & & $<0.9$\\
$\bar B^0\to\pi^0\rho^0$ & $ 0.54|V_{ub}|^2a_2^2$
& 0.092 & 0.14 &  &$<1.8$\\
$\bar B^0\to\rho^0\rho^0$ & $ 0.39|V_{ub}^2a_2^2$
& 0.067 & 0.05&  &$<28$\\
$\bar B^0\to D_s^-\pi^+$ & $0.285|V_{ub}|^2a_1^2$
& 0.85& 8.1 &1.9  &$<28$\\
$\bar B^0\to D_s^{*-}\pi^+$ & $1.06|V_{ub}|^2a_1^2$
& 3.1 & 6.1&2.7 & $<50$\\
$\bar B^0\to D_s^-\rho^+$ & $0.269|V_{ub}|^2a_1^2$
& 0.80 & 1.2 &1.0 & $<70$\\
$\bar B^0\to D_s^{*-}\rho^+$ & $2.25|V_{ub}|^2a_1^2$
& 6.7 & 4.5 &5.4 & $<80$\\
$B^-\to\pi^0\pi^-$ & $0.169|V_{ub}|^2(a_1+a_2)^2$
& 0.78 & 1.4 & & $<1.7$\\
$B^-\to\pi^0\rho^-$ & $0.438|V_{ub}|^2(a_1+0.52a_2)^2$
& 1.7 & 2.7 & & $<7.7$\\
$B^-\to\rho^0\pi^-$ & $0.120|V_{ub}|^2(a_1+1.95a_2)^2$
& 0.77 & 0.7& & $<4.3$\\
$B^-\to\rho^0\rho^-$ & $0.411|V_{ub}|^2(a_1+a_2)^2$
& 1.8 & 1.1& & $<100$\\
$B^-\to D_s^-\pi^0$ & $0.146|V_{ub}|^2a_1^2$
& 0.44& 3.9 &1.8  &$<20$\\
$B^-\to D_s^{*-}\pi^0$ & $0.545|V_{ub}|^2a_1^2$
& 1.6 & 3.0 &1.3 & $<33$\\
$B^-\to D_s^-\rho^0$ & $0.138|V_{ub}|^2a_1^2$
& 0.41 & 0.6&0.5 &$<40$\\
$B^-\to D_s^{*-}\rho^0$ & $1.16|V_{ub}|^2a_1^2$
& 3.4 & 2.4 &2.8 & $<50$\\
\hline
\hline
\end{tabular}
\end{table*}

We see that our results for the `class I' nonleptonic $B\to
D^{(*)}\pi(\rho)$ decays are close to the improved BSW model
predictions \cite{NRSX}, while our results for the `class II' and
$a_2$ contributions to `class III' decays are smaller than those of
\cite{NRSX}. These contributions come from $B\to\pi(\rho)$ transition
form factors, which have a different $q^2$ behaviour in our and the
BSW models. The BSW model assumes universal $q^2$ dependence of all
$B\to\pi(\rho)$ form factors. As already mentioned in our model we
find the $A_1$ and $F_0$ form factors to have a different $q^2$
dependence  than that of the other form factors 
\cite{semil,EFG}. The form factor $F_0$ in our
model decreases with the growing of $q^2$  in the
kinematical range of interest for energetic nonleptonic decays.
Note that our value for $B\to\pi$ form factors at $q^2=0$ is
approximately 1.5 times less than that of BSW, while the values for
$B\to\rho$ form factors are close in both models.

Our predictions for the branching ratios of $B\to D^{(*)}M$
nonleptonic decays presented in Table~1 agree with experimental data
within errors. Thus we can conclude that factorization works rather
well for `class I' and `class III' decays $B\to D^{(*)}\pi(\rho)$.
However, an improvement of experimental accuracy is needed to make a
definite conclusion. It will be very interesting to measure the `class
II' decay $\bar B^0\to D^{(*)0}\pi(\rho)^0$ branching ratios. Such
measurement will be the test of factorization for `class II'
nonleptonic decays and will help to constrain the  $q^2$
dependence of $B\to\pi(\rho)$ form factors.

We also present in Table~1 the predictions for $B\to D^{(*)}D_s^{(*)}$
nonleptonic decays, where only heavy mesons are present in the final
state. The factorization is less justified for such decays. However,
as we see from  Table~1  our predictions, based on the
factorization, are consistent with the experimental data for these
decays too.

For the branching ratios of $B\to\pi(\rho)M$ nonleptonic decays
presented in Table~2 there are only experimental upper limits at
present. The measurement of these decays will allow the determination
of the CKM matrix element $|V_{ub}|$, which is poorly known. The
closest experimental upper limit to the theoretical predictions is for
the decay $\bar B^0\to\pi^+\pi^-$ \cite{ag}.
It is approximately 1.5 times
larger than our prediction and is very close to the result of
\cite{Dean}. From this upper limit on  $B(\bar B^0\to\pi^+\pi^-)$
we get the limit on $|V_{ub}|$ in our model
$$ |V_{ub}|<6.4\times 10^{-3},$$
which is close to the value previously found from
semileptonic $B\to\pi(\rho)l\nu$ decays \cite{semil}:
$$|V_{ub}|=(5.2\pm1.3\pm0.5)\times 10^{-3}.$$

The overall agreement of our predictions for two-body nonleptonic
decays of $B$ mesons with the existing experimental data
\cite{BHP,PDG} shows that the factorization approximation works
sufficiently well in the framework of our model. From another side, if
the factorization hypothesis is taken for granted, the  aforementioned
agreement confirms the selfconsistency of our approach, which
incorporates our previously obtained results for semileptonic and
leptonic decays of heavy mesons. In particular it would be quite
intersting to test the specific $q^2$ behaviour of the heavy-to-light
transition form factors $F_0$ and $A_1$ predicted by our model.

\end{document}